\documentclass[aps, floatfix, showpacs, superscriptaddress, preprint, pre]{revtex4-1}

\usepackage{amssymb,amsmath,graphicx}
\usepackage{placeins}
\usepackage{xcolor}

\newcommand{\eps}{\varepsilon}
\newcommand{\SUM}{\sum\limits}
\newcommand{\expp}{\operatorname{e}}
\newcommand{\INT}{\int\limits}

\newcommand{\E}{\mathbb{E}}

\begin{document}

\title{Coherence resonance in neuronal populations: mean-field versus network model}

\author{Emre Baspinar}
\email[]{emre.baspinar@inria.fr}
\affiliation{Inria Sophia Antipolis M\'{e}diterran\'{e}e Research Centre, 2004 Route des Lucioles, 06902 Valbonne, France}

\author{Leonhard Sch\"ulen}
\email[]{l.schuelen@campus.tu-berlin.de}
\affiliation{Institut f\"ur Theoretische Physik, Technische Universit\"at Berlin, Hardenbergstraße 36, 10623 Berlin, Germany}

\author{Simona Olmi}
\email[]{simona.olmi@fi.isc.cnr.it}
\affiliation{Inria Sophia Antipolis M\'{e}diterran\'{e}e Research Centre, 2004 Route des Lucioles, 06902 Valbonne, France}
\affiliation{CNR - Consiglio Nazionale delle Ricerche - Istituto dei Sistemi Complessi, 50019, Sesto Fiorentino, Italy}

\author{Anna Zakharova}
\email[]{anna.zakharova@tu-berlin.de}
\affiliation{Institut f\"ur Theoretische Physik, Technische Universit\"at Berlin, Hardenbergstraße 36, 10623 Berlin, Germany}

\begin{abstract}
The counter-intuitive phenomenon of coherence resonance describes a non-monotonic behavior of the regularity of noise-induced oscillations in the excitable regime, leading to
an optimal response in terms of regularity of the excited oscillations for an intermediate noise intensity. We study this phenomenon in populations of FitzHugh-Nagumo (FHN) neurons with different coupling architectures. For networks of FHN systems in excitable regime, coherence resonance has been previously analyzed numerically. Here we focus on an analytical approach studying the mean-field limits of the locally and globally coupled populations. The mean-field limit refers to the averaged behavior of a complex network as the number of elements goes to infinity. We derive a mean-field limit approximating the locally coupled FHN network with low noise intensities. Further, we apply mean-field approach to the globally coupled FHN network. We compare the results of the mean-field and network frameworks for coherence resonance and find a good agreement in the globally coupled case, where the correspondence between the two approaches is sufficiently good to capture the emergence of anticoherence resonance. Finally, we study the effects of the coupling strength and noise intensity on coherence resonance for both the network and the mean-field model.
\end{abstract}




\maketitle

\section{Introduction}

All real-world processes are affected by random fluctuations that are intrinsically produced by the system itself and/or introduced by the extrinsic mechanisms to the system. The fluctuations are modeled mathematically as noise \cite{STR65_67,ARN98, ANI07}. In neural networks noise occurs naturally, for example, due to random synaptic input from other neurons, spontaneous neural activity or random opening and closing of ionic channels resulting in so-called intrinsic brain noise \cite{DEC09a,PIS14,RUN16}. Investigation of noise impact in the brain, and in nonlinear dynamical networks in general, is a challenging problem. Noise can give rise to new dynamic behavior, e.g., stochastic bifurcations \cite{ARN98,ZAK10a,ZAK11} or stochastic synchronization \cite{ANI97,ZAK13}. It can even induce partial synchronization patterns such as chimera states \cite{SEM16,ZAK17,ZAK17a, olmi2019chimera, ZAK20}. Intriguingly, random fluctuations do not always have a destructive impact deteriorating the regularity of a deterministic system. Instead they can play a constructive role and lead to an increase of coherence for increasing noise intensities. The most prominent examples are stochastic resonance \cite{BEN82, douglass1993noise, moss1994stochastic, wiesenfeld1995stochastic, gingl1995non, rappel1996noise, bulsara1996threshold, collins1996noise, GAM98, ANI99, muratov2005self, MCD09, GAM09, SEM15b} that is observed for periodically driven bistable systems and \textit{coherence resonance} that takes place in autonomous systems, i.e., under purely noise excitation without any periodic driving signal \cite{GAN93,PIK97,NEI97, lee1998coherence, pradines1999coherence, giacomelli2000, miyakawa2002experimental, tateno2004, LIN04, DEV05,USH05,ZAK10a, GEF14,SEM15,JUS16}. 

The counter-intuitive phenomenon of coherence resonance was originally detected for an excitable FitzHugh-Nagumo neuron \cite{PIK97}. It describes a non-monotonic behavior of the regularity of noise-induced oscillations in the excitable regime, leading to an optimal response in terms of regularity of the excited oscillations for an intermediate noise intensity. 
Since the discovery of coherence resonance, it has been investigated theoretically and experimentally in various systems and networks \cite{HU00,BAL04,SCH04b,LIN04,HAU06,ZIE13,BAL14,JUS16,SEM16,ZAK17a,MAS17}. It has been shown that coherence resonance can be observed not only in excitable \cite{PIK97, AUS09}, but also in non-excitable systems \cite{USH05,ZAK10a,ZAK11,ZAK13,GEF14,SEM15}. Depending on the nature of the external noisy inputs, different mechanisms for coherence resonance have been observed, like the double coherence resonance, occurring for an optimal combination of noise variance and correlation of inputs stimulating a single FitzHugh-Nagumo neuron \cite{kreuz2006double, kreuz2007coherence}. On the other hand, in complex networks of FitzHugh-Nagumo units, the existence of coherence resonance has been reported for one-layer \cite{MAS17} and two-layer \cite{SEM18} networks. Further topologies include local, nonlocal, global coupling, lattice networks as well as more complex structures such as random or small-world networks \cite{WAN00,KWO02,SUN08,YIL16,MAS17,AND18}. 
 
Interestingly, large ensembles of neurons in the brain demonstrate a rich variety of coherent dynamics at the macroscopic scale which results from random perturbations \cite{brunel2003determines}. Therefore, understanding coherence resonance is important for the study of brain and neural networks. Coherence is significant for communication between brain regions \cite{DEC16} and, as recently suggested in \cite{PIS19}, the improvement of neural communication can be reached via coherence resonance. In more detail, the brain adjusts its internal noise in order to maximize the coherence.
Futhermore, information processing and its encoding for transmission to different areas of the brain require a coherent activity of neuronal populations. In particular, information processing in the brain can be represented as a non-stationary spatiotemporal process of activity propagation \cite{Kirst2016, Palmigiano2017}. In this view, brain activity during task conditions simultaneously evolves in a hierarchy of characteristic network activations. For instance, information processing in sensorimotor coordination \cite{Daffertshofer} and auditory, visual and linguistic tasks \cite{Barry_Horwitz} show robust propagation through well-tuned activation chains of characteristic subnetworks.
Especially in the sensory cortex, many neurons locally sensitive to similar stimulus features give a similar response to a given input stimulus (see for example \cite{hubel1959receptive, hubel1962receptive, averbeck2006neural} for neurophysiological studies, and also \cite{citti2006cortical, sarti2008symplectic, baspinar2018geometric, baspinar2019sub} for some models using the local tuning feature of neurons). This suggests that the activity of such neurons can be measured and studied at a macroscopic scale, which provides reliable data due to the averaging effects diminishing the independent chaotic random behaviors of single neurons observed at a microscopic scale.

The growing interest in the phenomenon of coherence resonance for neural networks is confirmed by a number of works \cite{WAN00,KWO02,TOR03,DU07,SUN08,YIL16,MAS17}, including very recent ones \cite{SEM18,AND18,PIS19}. 
While the majority of these investigations is based on numerical simulations or on experimental data, much less attention is paid to the analytical treatment of coherence resonance in complex networks. Although analytical treatment has been provided for single systems (e.g., for FHN system in \cite{LIN99a}, for generalized van der Pol oscillator in \cite{ZAK13} and for leaky integrate-and fire neuron in \cite{olmi2017exact}), it remains a demanding problem for networks. Here we address this challenging question by developing the mean-field framework for analyzing coherence resonance in neural populations. Using a paradigmatic model of FHN neuron, we study the phenomenon of coherence resonance both at the network level (i.e., where we model each neuron in the population as a perturbed coupled FHN system) and at the level of \emph{mean-field limit} (known also as \emph{thermodynamic limit}), which is the asymptotic limit of the averaged network as we send the number of neurons to infinity.

Our main contributions are at both theoretical and numerical levels. At the theoretical level, we provide a mathematical underpinning of the results obtained from numerical simulations of locally and globally coupled neural networks presented in \cite{MAS17}. In particular we derive a mean-field description for the locally coupled case which coincides with the network framework for small noise intensities and low excitability threshold. 
For the globally coupled case, we employ the mean-field description provided in \cite{baladron2012mean,bossy2015clarification} for a generic family of stochastic differential equations of stochastic FHN type and we decline the model in two versions, which differ from the number of employed noise terms: a single additive noise term in the differential equation for the recovery variable or two different noise terms, one for each variable equation. The latter mean-field model, which represents an extended version of the former, has been designed to take into account 
the noise effects experienced by electrical synapses when arranging the passive flow of the ionic current through the pores between neurons: this flow, together with the charge carriers, has
a stochastic nature.

At the numerical level, we investigate the emergence of anticoherence resonance in the globally coupled framework, which results in different outcomes depending on the investigated model.
When one single noise term is present, anticoherence resonance does not emerge in the system, while, when two noise terms are present, anticoherence resonance emerges for large noise intensity and it is characterized by an alteration of the dynamics, which is now guided by the noise.

The manuscript is organized as follows. In Section \ref{sec:network_equations}, we provide the network settings for each topological architecture. Afterwards, we explain the associated mean-field frameworks in Section \ref{sec:meanFieldPopulation}. In Section \ref{sec:simulations_results}, we explain the experimental setting and present our simulation results where we compare the coherence resonance results obtained from the mean-field systems and network framework to find out to which level they provide the same outcome, i.e., we identify the limitations of the mean-field approach. Finally, we give the conclusions in Section \ref{sec:conclusio}.


\section{Network equation}
\label{sec:network_equations}

Several dynamical models have been proposed to study both single neuron and coupled neuronal population behaviors such as the Hodgkin-Huxley model \cite{hodgkin1952quantitative}, or other conductance-based models, like the Morris-Lecar model \cite{morris1981voltage}.
Conductance-based models are the simplest possible biophysical representation of an excitable cell, such as a neuron, in which its protein molecule ion channels are represented by conductances and its lipid bilayer by a capacitor. However, other types of models have been developed, that predict the dynamics of the membrane output voltage as a function of electrical stimulation at the input stage,
like Hindmarsh and Rose \cite{hindmarsh1984model}, integrate-and-fire \cite{gerstner2002spiking, brette2007simulation, izhikevich2007dynamical} or Galves-L\"{o}cherbach model \cite{galves2013infinite}.
At the mean-field level, heuristic firing rate models are commonly used, like the Wilson-Cowan excitatory-inhibitory neural mass model \cite{wilson1972excitatory,wilson1973mathematical} and its stochastically modified versions \cite{touboul2012noise}. Only recently have been developed neural mass models that are not derived heuristically, but that reproduce exactly the dynamics of excitatory and inhibitory networks of spiking neurons for any degree of synchronization \cite{montbrio2015macroscopic, devalle2017firing, ceni2020cross, segneri2020theta}. In particular these neural masses reproduce the macroscopic dynamics of quadratic integrate-and-fire neurons, which are normal forms for the saddle-node on a limit cycle bifurcation (SNIC) \cite{ermentrout1986parabolic} and describe, in general, the dynamics of class I neurons (i.e. neurons with a continuous gain function), to which belong the Hindmarsh-Rose model and the Morris-Lecar model under some circumstances. 
In the present work, we are particularly interested in the celebrated FitzHugh-Nagumo (FHN) model \cite{fitzhugh1961impulses, nagumo1962active} which represents a reduced version of the Hodgkin–Huxley model. It still captures closely the dynamical behaviors produced by the Hodgkin–Huxley model and has the advantage of facilitating efficient large-scale simulation of groups of neurons.

Network equations describe the dynamics of each neuron belonging to the network at a microscopic level. The classical deterministic Fitzhugh-Nagumo (FHN) equations describing the evolution of a single neuron belonging to a coupled population read as
\begin{equation}\label{eq:deterministic_ith_neuron_coupled_eqns}
\begin{split}
\varepsilon \frac{d u_i(t)}{dt}=  & f(u_i(t), v_i(t))+\frac{\sigma}{2P}\SUM_{j=i-P}^{i+P}[u_j(t)-u_i(t)]\\
 = & u_i(t)-\frac{u_i(t)^3}{3}-v_i(t) +\frac{\sigma}{2P}\SUM_{j=i-P}^{i+P}[u_j(t)-u_i(t)],\\
\frac{dv_i(t)}{dt}= & g_a(u_i(t))=u_i(t)+a,
\end{split}
\end{equation}
where $\sigma$ is a positive constant denoting the coupling strength.
$u_i$ and $v_i$ represent the activator (membrane potential) and inhibitor (recovery) variables of the  $i$-th neuron respectively ($i=1,\ldots, N$, where $N$ is the population size). $\varepsilon>0$
is responsible for the time scale separation of fast activator and slow inhibitor, being a small parameter. Here, we fix $\varepsilon=0.01$. Parameter $a$ determines the nature of the equilibrium points and thus the excitability threshold of the isolated, uncoupled neuron. In particular, the parameter $a$ serves as a threshold in our model and determines whether the single neuron is in the excitable $|a|>1$ or in the oscillatory $|a|<1$ regime. The single neuron undergoes a Hopf bifurcation when $a = 1$.
Finally $P$ denotes, for the $i$-th neuron, the number of its nearest neighbours in each direction of the ring. Thus, it determines the topology of the population: if $P=1$ we have local coupling, if $P=(N-1)/2$ (we assume $N$ is odd) we have global coupling and finally if $1<P<(N-1)/2$, we have nonlocal coupling. 

We assign a stochastic behavior to the system, by adding a Gaussian white noise term $dW_i(t)$ to the recovery variable equation of each neuron, as described in \cite{MAS17} and \cite{SEM18}. The noise term $dW_i(t)$ is built, for each neuron $i$, from an independent Wiener process. More precisely
\begin{equation}\label{eq:noiseProperties}
\E[dW_i(t)]=0,\quad \E[dW_i(t)dW_j(t^\prime)]=\delta\Big ( (i-j)(t-t^\prime) \Big ), \forall i,j
\end{equation}
with $\mathbb{E}$ and $\delta$ denoting expectation value and Dirac delta function, respectively. For the sake of simplicity, we drop showing the time dependency explicitly from now on, as long as the otherwise is required. The stochastic FHN equations read as follows: 
\begin{equation}\label{eq:stochNetwork}
\begin{split}
\varepsilon du_i(t)= & f(u_i(t), v_i(t))dt+\frac{\sigma}{2P}\SUM_{j=i-P}^{i+P}[u_j(t)-u_i(t)]dt,\\
dv_i(t)= & g_a(u_i(t))dt+ \sqrt{2D}\, dW_i(t),  \qquad i=1,\dots,N,
\end{split}
\end{equation}
where $D$ denotes the level of noise intensity. 
Gaussian white noise is intended to represent triggering perturbations that alters the state of the system. Here we focus on the dynamics of neuronal populations in their excitable regime. In this regime, a sufficiently strong perturbation triggers the whole population to produce spikes before the population comes back to its steady state. If the perturbation is not strong enough, the population relaxes back to its unique stable steady state without producing a spike. In particular, we are interested in the state of the network characterized by the best temporal regularity of the noise-induced spiking dynamics achieved for an intermediate optimal noise intensity, i.e., when the network undergoes coherence resonance. In Sec. IV, when comparing the network dynamics emergent in the network with the mean-field prediction, in the globally coupled regime, we will also integrate a second set of equations that differ from Eqs. \eqref{eq:stochNetwork} by adding a second noise term $\sqrt{2\varepsilon \bar{D}}\, d\bar{W}_i(t)$ in the first differential equation, where $\bar{D}$ denotes the level of noise intensity and $d\bar{W}_i(t)$ represents, as before, a Gaussian noise source. The reason for this will become clear when introducing the extended version of the mean-field limit associated to the globally coupled FHN system (see Eqs. \eqref{eq:stochFHN} in Sec. III).

\subsection{Macroscopic indicators of coherence resonance}
Coherence resonance characterizes the emergence of relatively coherent noise-induced oscillations occurring for an optimal noise intensity. It was initially found in a single FHN system in the excitable regime and later detected for neural networks.
There exist several different measures for quantifying coherence resonance, such as the normalized standard deviation of the interspike interval, the correlation time, and the signal-to-noise-ratio \cite{GAN93,PIK97,GEF14}. Since in the present work we deal with a neural model showing spiking behaviour, it is convenient to use the standard deviation of the interspike interval (ISI),
defined as
\begin{equation}
\label{stdev_ISI}
 R_{ISI}=\displaystyle\frac{\sqrt{\langle{{t_{ISI}^2}}\rangle - {\langle{t_{ISI}}\rangle}^2}}	{\langle{t_{ISI}} \rangle},
\end{equation}
where $t_{ISI}$ is the time between two subsequent spikes and $\langle \cdots \rangle$ indicates the average over the time series. A system undergoing coherence resonance will show a pronounced minimum in the value of $R_{ISI}$ \cite{PIK97}. The definition \eqref{stdev_ISI} is limited to characterizing coherence resonance for a single FitzHugh-Nagumo oscillator. For a network of oscillators, coherence resonance can be measured by redefining $R$ as follows \cite{MAS17}:  
\begin{equation}
R=\frac{\sqrt{\langle \overline{{t_{ISI}^2}}\rangle - {\langle \overline{t_{ISI}}\rangle}^2}}	{\langle\overline{t_{ISI}} \rangle},
\label{eq:R_network}
\end{equation}
where the over-bar indicates the additional average over nodes. 

\begin{figure}
	\includegraphics[width=\textwidth]{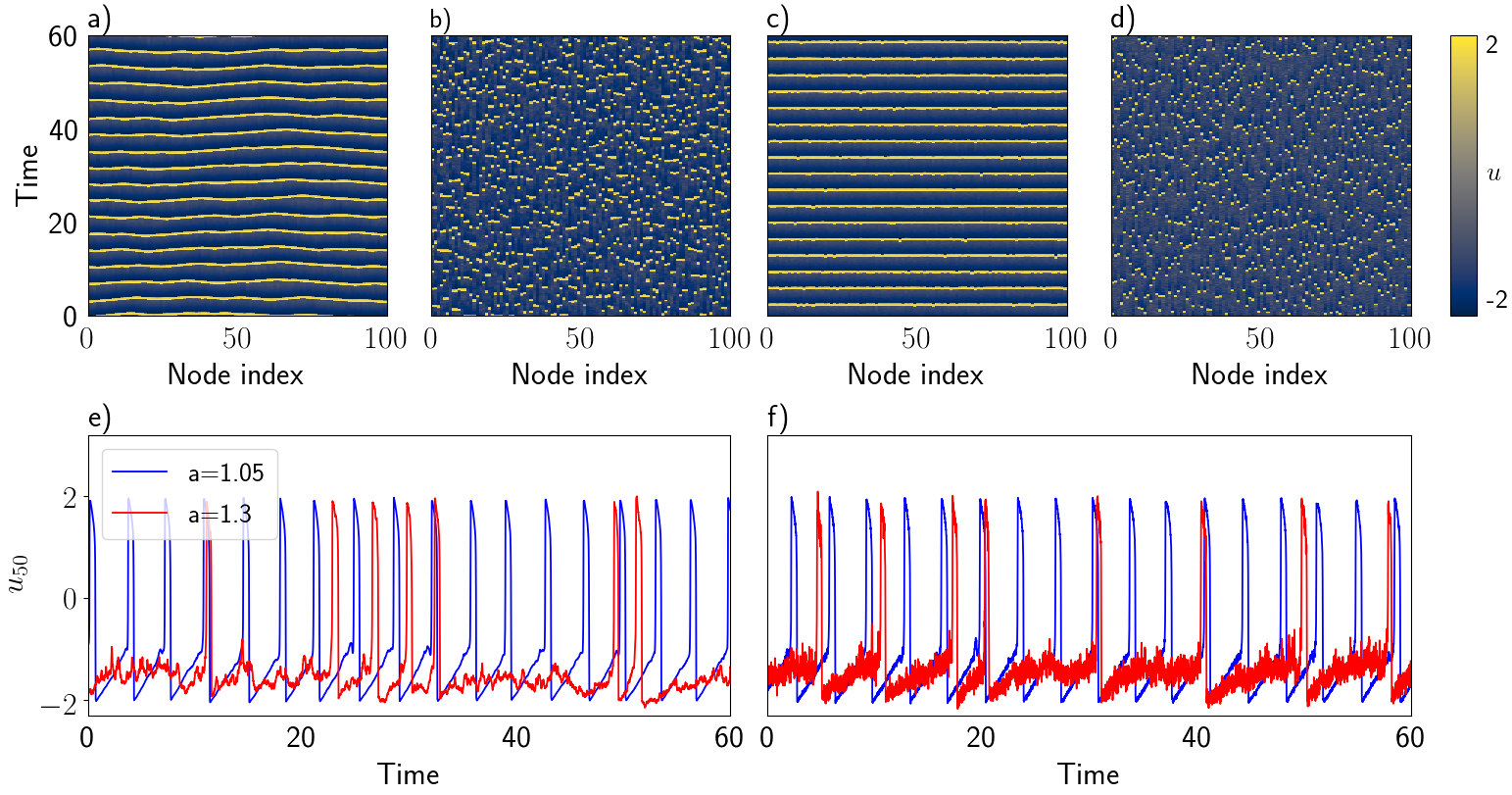}
	\caption{Space-time plots for optimal $D$ values (i.e., $R$ takes minimum values) for different network topology and $a$ values: a) a locally coupled network with $a=1.05$ and $D=0.001$, b) a locally coupled network with $a=1.3$ and $D=0.08$, c) a globally coupled network with $a=1.05$ and $D=0.0008$, d) a globally coupled network with $a=1.3$ and $D=0.0158$. Time series of one selected neuron in a locally (panel e) and globally (panel f) coupled network for $a=1.05$ (blue) and $a=1.3$ (red). Other parameters: $N=100$, $\varepsilon=0.01$, $\sigma=0.1$.}
	\label{fig:coher_example}
\end{figure}

To illustrate the dynamics of the system in the regime of coherence resonance we provide here space-time plots and time series for a locally and globally coupled network in the stochastic regime (see Fig. \ref{fig:coher_example}). In more detail, we show the dynamical behavior emerging for different threshold parameter values ($a=1.05$ and $a=1.3$) for the locally coupled (Fig. \ref{fig:coher_example} a) 
and b)) and the globally coupled (Fig. \ref{fig:coher_example} c) and d)) case. 
In the locally coupled case we observe, for the lower threshold value ($a=1.05$), coherent response among all network elements that spike regularly and nearly at the same time (panel a). The case of higher threshold $a=1.3$ shows a more irregular response, while still having a rather small $R$ value (panel b). When the threshold is increased, it is harder for the system to overcome it. 
A similar scenario emerges when passing from local to global coupling. For the lower threshold value the coherent response of the network is clearly observable. In this case the FHN neurons spike in a highly synchronized way (see the straight yellow lines in panel c), due to the fact that each element receives the input from the entire network being connected with all the others. The simultaneous interaction among all neurons enhance them to overcome the threshold all at once. At the contrary, in the locally coupled case, each neuron pulls only its immediate neighbors over the threshold, which explains why the excitation needs some time to ``travel'' over the ring. Similarly to what shown in panel b), the case of higher threshold value gives a more irregular picture even in the globally coupled configuration (panel d). The same conclusions can be drawn looking at the time series of single units. In particular in Fig. \ref{fig:coher_example} e) are shown the time series of one selected neuron in the locally coupled network for $a=1.05$ (blue curve) and $a=1.3$ (red curve), while Fig. \ref{fig:coher_example} f) illustrates the time series of the same element in the globally coupled case for $a=1.05$ (blue curve) and $a=1.3$ (red curve). The higher temporal regularity for the lower threshold parameter $a$ can be clearly seen for both coupling topologies.

\section{Mean-field population equations}
\label{sec:meanFieldPopulation}

\subsection{Globally coupled equations}

In the globally coupled mean-field framework, we have $P=N/2$ as $N\rightarrow\infty$. We assume that the coupling strength is the same and constant for all neurons in a single population. 
To avoid using negative indices in Eqs. \eqref{eq:stochNetwork} we change the indexing of the coupling term according to
\begin{equation}
\SUM_{j=1}^{N+1}[u_i-u_j],
\end{equation}
without losing the generality. 

We emphasize that the noise terms in Eqs.~\eqref{eq:stochNetwork} are assumed to be independent and identically distributed for each neuron and also for each state variable. This allows us to describe, in the thermodynamic limit, each state variable of a neuron as a continuous set of random variables, each one corresponding to an instant of time. In other words, each state variable can be thought of as a stochastic process representing the values of the state variable changing randomly in time. As such, the state variables $u_i$ and $v_i$ of the $i$-th neuron (for any $i=1\dots, N$) evolve in accordance with their associated probability distributions, which converge in the thermodynamic limit, to the ones of the mean-field state variables $u$ and $v$ characterized by
\begin{equation}
\lim_{N\rightarrow \infty}\;\max_{i=1,\dots,N}\mathbb{E}\Big [ \sup_{s\leq t}\Big( u_i(s)-u(s) \Big ) ]^2 \Big ]=0,\quad\quad
\lim_{N\rightarrow \infty}\;\max_{i=1,\dots,N}\mathbb{E}\Big [ \sup_{s\leq t}\Big( v_i(s)-v(s) \Big ) ]^2 \Big ]=0.
\end{equation}
As it was shown in \cite{baladron2012mean, bossy2015clarification}, the state variables $u$ and $v$ comprise the solution to the following mean-field system: 
\begin{equation}\label{eq:globallyCopuledMF}
\begin{split}
\varepsilon du= & f(u,v)dt + \sigma\big(\E[u] -u \big)dt,\\
dv= & g_a(u)dt+\sqrt{2D}\, dW,
\end{split}
\end{equation}
where $dW$ denotes Gaussian white noise with the properties given in Eqs. \eqref{eq:noiseProperties} and $D$ defines the noise intensity level. Here $\sigma$ represents the coupling constant, as introduced in the network Eqs. \eqref{eq:stochNetwork}. This mean-field equation was derived in \cite{baladron2012mean,bossy2015clarification} from a network of FHN neurons of the same type as Eqs.~\eqref{eq:stochNetwork}, based on two steps: i) the first step shows that a unique solution to Eqs.~\eqref{eq:globallyCopuledMF} exists for a finite time, under the assumption that the terms  $f$ and $g_a$ are locally sufficiently regular (locally Lipschitz); ii) the second step shows that for each neuron the probability distributions of the processes $u_i$ and $v_i$ converge towards the probability distributions of the mean-field state variables $u$ and $v$, respectively.

While the mean-field framework shown in Eqs.~\eqref{eq:globallyCopuledMF} corresponds to the classical mean-field limit of the globally coupled networks of FHN oscillators of the type given in \cite{MAS17,  SEM18}, we consider in the following  an extended version of the mean-field limit associated to the globally coupled FHN system. 
In \cite{baladron2012mean,bossy2015clarification} it was shown that the same results and properties as in Eqs.~\eqref{eq:globallyCopuledMF} hold also for the extended version, which is adapted from \cite{baladron2012mean, bossy2015clarification}, as follows: 
\begin{equation}
\label{eq:stochFHN}
\begin{split}
\eps\, du= & f(u , v)dt + \sigma (\mathbb{E}[u]-u)dt+\sqrt{2\eps\bar{D}} \,d \bar{W},\\
d v= & g_a(u)dt+\sqrt{2D} \, d W,
\end{split}
\end{equation}
where $d\bar{W}$, $dW$ are the noise terms with standard normal distribution, constructed from two independent Brownian motions $\bar{W}$ and $W$. We denote the noise intensity levels as $\bar{D}$, $D$. 
As for Eqs. \eqref{eq:globallyCopuledMF}, the presence of $\sigma(\mathbb{E}[u]-u)dt$ requires studying the state variable $u$, whose solution depends on its own expectation value.

The choice of this extended model is motivated by the fact that the coupling term is based on modeling electrical synapses. Such synapses arrange the passive flow of the ionic current through the pores between neurons and, as a result of the stochastic nature of ionic currents and charge carriers, they experience noise effects. The coupling related terms persist in the mean-field framework of the globally coupled configuration, while this is not the case in the locally coupled case (see Eqs.~\eqref{eq:FHN_local_coupled}). Therefore, we model the noise effects arising from the global coupling by introducing an additional white noise term ($d\bar{W}$) in Eqs.~\eqref{eq:globallyCopuledMF}, as shown in Eqs.~\eqref{eq:stochFHN}. In order to guarantee that $d\bar{W}$ does not dominate $dW$, neither the whole system behavior, we impose $\bar{D}$ to be of the same order as $D$ and the scaling $\sqrt{\eps}$. Finally, we remark that, when comparing the results obtained from the simulation of Eqs. \eqref{eq:stochFHN} with those obtained from the network, the corresponding network equations are described, for consistency, by Eqs. \eqref{eq:stochNetwork} with an additional noise term, $\displaystyle\sqrt{2\eps\bar{D}}\, d\bar{W}_i(t)$, in the first equation, where $\bar{D}$ denotes the level of noise intensity for the Gaussian white noise $d\bar{W}_i(t)$. 

The mean-field limit models the population behavior by employing a single FHN system once the number of the neurons in the population is sufficiently high, whereas, for the network equations, we need a separate FHN equations system for each neuron in the population. In other words, the network equations require a high-dimensional dynamical system while the mean-field limit requires only a two-dimensional dynamical system, being a good representative of the averaged dynamics of the population and making analytical treatment of the system feasible. However, studying the statistics of such a system is not trivial since the right-hand side requires the expectation of the solution of the equation. Yet, it is possible to use a semi-analytical approach, where we obtain the first-moment statistics of $u$ from numerical simulations of the network equations given by Eqs. \eqref{eq:stochNetwork}, as shown in \cite{baladron2012mean}. Practically, the expectation value of $u$ will be introduced numerically in the differential equations \eqref{eq:stochFHN}, by obtaining the statistics of the state variables from the network simulations. In this way it will be possible to investigate the interplay of coupling and noise to determine the emergence of coherence resonance and to provide an analytical framework for the numerical results. 

Finally, we note that, although the notation for the mean-field state variables $u$ and $v$ will be the same for both globally and locally coupled settings, the corresponding definitions are different and, in the rest of the paper, it should be tacitly understood from the coupling type.

\subsection{Locally coupled equations}
In the locally coupled topology $P=1$ and the network equations for the $i$-th neuron can be written as
\begin{equation}\label{eq:LocalSingleNeuron}
\begin{split}
\varepsilon du_i= & f(u_i, v_i)dt+\frac{\sigma}{2}\Big( u_{i-1}+u_{i+1}-2u_i \Big )dt,\\
dv_i= & g_a(u_i)dt+ \sqrt{2D}\,dW_i,\quad i=1,\dots,N.
\end{split}
\end{equation}
We assume to have periodic boundary conditions: the $i$-th neuron is coupled to the $(i-1)$-th and $(i+1)$-th neurons for all $i\in \{1,2,\dots,N\}$, while it holds that
\begin{equation}\label{eq:ring_Couipling}
u_{N+K}=u_K\quad \text{for all}\;K\in \{-N,-N+1,\dots,N-1,N\}.
\end{equation}
The average dynamics of a locally coupled system in the thermodynamic limit can be found by considering that
\begin{equation}\label{eq:averaging_LocalCase}
\begin{split}
\frac{\varepsilon}{N}\SUM_{i=1}^N  du_i= &\frac{1}{N}\SUM_{i=1}^N f(u_i, v_i)dt+\frac{\sigma}{2N}\SUM_{i=1}^N\Big( u_{i-1}+u_{i+1} - 2 u_i\Big)dt,\\
\frac{1}{N}\SUM_{i=1}^N dv_i= & \frac{1}{N}\SUM_{i=1}^N g_a(u_i)dt+ \frac{1}{N}\SUM_{i=1}^N\sqrt{2D}\,dW_i.
\end{split}
\end{equation}
where the term proportional to the coupling constant $\sigma$ in the r.h.s. of the first equation vanishes for $N\rightarrow\infty$. Moreover, we define in the following
\begin{equation}
u:=\frac{1}{N}\SUM_{i=1}^N u_i,\quad\quad v:=\frac{1}{N}\SUM_{i=1}^N v_i,
\end{equation}
and
\begin{equation}\label{eq:centralLimit}
\frac{1}{N}\SUM_{i=1}^NdW_i=dW,
\end{equation}
where $dW$ is a Gaussian white noise as a result of the central limit theorem. It is not straightforward to write the average dynamics directly from Eqs. \eqref{eq:averaging_LocalCase} (neither the exact mean-field limit) due to the nonlinearity of $f$. In order to handle the nonlinearity, we approximate the state variables $u_i$ as random variables distributed according to a Gaussian distribution, as described in \cite{buric2010mean} (see also \cite{rodriguez1996statistical, tuckwell1998hc, tanabe2001dynamics, zaks2005noise} for details regarding the use of Gaussian random variables in such approximations). This approximation assumes that the excitable system is sufficiently close to the equilibrium point and the noise intensity $D$ is small. We employ the law of large numbers (see, for example \cite{Etemadi1981}), more precisely
\begin{equation}\label{eq:largeNumber}
\frac{1}{N}\SUM_{i=1}^N u_i=\mathbb{E}[u_i]\quad \text{as}\quad N\rightarrow \infty,
\end{equation}
and write the mean-field limit from the average dynamics given in Eqs. \eqref{eq:averaging_LocalCase}. Since we approximate $u_i$ as a Gaussian random variable with expectation value $u$ and variance $\rho^2$, we have
\begin{equation}\label{eq:assumptionsFromGaussian}
\begin{split}
\mathbb{E}[u_i]= & \frac{1}{\sqrt{2\pi \rho^2}}\INT_{-\infty}^\infty u_i\expp^{\frac{-(u_i-u)^2}{2\rho^2}}du_i=u,\\
\mathbb{E}[u_i^3]= & \frac{1}{\sqrt{2\pi \rho^2}}\INT_{-\infty}^\infty u_i^3\expp^{\frac{-(u_i-u)^2}{2\rho^2}}du_i=u^3+3\rho^2u.
\end{split}
\end{equation}
By implementing Eqs. \eqref{eq:centralLimit}, \eqref{eq:largeNumber} and \eqref{eq:assumptionsFromGaussian} in Eqs. \eqref{eq:averaging_LocalCase} we find, in the limit $N\rightarrow \infty$, that
\begin{equation}
\begin{split}
\varepsilon du= & f(u,v)dt-\rho^2 u\,dt,\\
dv= & g_a(u)+\sqrt{2D}\,dW,
\end{split}
\end{equation}
and more explicitly
\begin{equation}\label{eq:FHN_local_coupled}
\begin{split}
\begin{split}
\varepsilon du= &\Big( (1-\rho^2)u-\frac{u^3}{3}-v\Big) dt,\\
dv= & \Big(u+a\Big)dt+\sqrt{2D}\,dW.
\end{split}
\end{split}
\end{equation}
The fact that the coupling term vanishes in Eqs. \eqref{eq:averaging_LocalCase} indicates that there is no effect of the coupling at the mean-field level.
Moreover, since we do not have any information about $\rho$ a priori, we obtain the $\rho$ values from the numerical network simulations and introduce them at each time
sample when we perform the numerical integration of Eqs. \eqref{eq:FHN_local_coupled}. 

\subsection{Nonlocally coupled equations}
Finally we consider the intermediate nonlocally coupled case with $1<P<\frac{N}{2}$.
It is convenient to distinguish between two cases, that represent two classes of systems \cite{golomb2001mechanisms}: sparse (or strongly diluted) networks, where $P\ll N$, and specifically $P$ is independent of $N$ as $N\rightarrow \infty$; massive networks, where $P$ is proportional to the network size $N$. In our ring topology, these cases can be translated in the following limits
\begin{equation}
\lim_{N\rightarrow \infty}\frac{P}{N}=0\quad\text{and}\quad \lim_{N\rightarrow \infty}\frac{P}{N}=C,
\end{equation}
where $C$ is a constant value and $C\leq 1/2$ for the definition of ring topology. 
In the first case ($P\ll N$), if we fix $P$ to be a constant connectivity, we can write the following averaged network equations for $N\rightarrow \infty$ 
\begin{equation}
\begin{split}
\frac{\varepsilon}{N}\SUM_{i=1}^N du_i= &\frac{1}{N}\SUM_{i=1}^N f(u_i, v_i) dt+\frac{1}{N}\SUM_{i=1}^N\frac{\sigma}{2P}\Big ( u_{i-P}+\dots +u_{i+P} - 2P u_i\Big )dt,\\
\frac{1}{N}\SUM_{i=1}^N dv_i= & \frac{1}{N}\SUM_{i=1}^Ng_a(u_i)dt+\frac{1}{N}\SUM_{i=1}^N \sqrt{2D}\,dW_i,
\end{split}
\end{equation}
which turns out to be the same as the locally coupled system given in Eqs. \eqref{eq:FHN_local_coupled}. In the second case ($P\propto N$), we may write $P$ as a function of $N$, i.e. $P=P(N)$,
such that the same limit holds for $P$ and $N$, when $N\rightarrow \infty$. This means that, in the thermodynamic limit, the mean-field system will converge to the globally coupled case given in Eqs. \eqref{eq:globallyCopuledMF} as long as the coupling constant $\sigma$ is rescaled in accordance with the limit value comparable to $C$.

We conclude that nonlocally coupled topology cannot be treated as a separate one, since its dynamics can be attributable to the one emergent in either sparse or massive networks.
This case confirms what already found in literature: i) for massive networks, i.e. when the connectivity scales with $N$, the network behaves like a globally coupled system with a rescaled coupling constant to account for the different fraction of active links \cite{olmi2010collective}; ii) for sparse networks, characterized by constant connectivity, not increasing with $N$, the thermodynamic limit shows a completely different behavior, typical of locally coupled topology \cite{luccioli2012collective}.

\section{Simulation results}
\label{sec:simulations_results}

\subsection{Globally coupled framework}

\begin{figure*}
	\includegraphics[width=\textwidth]{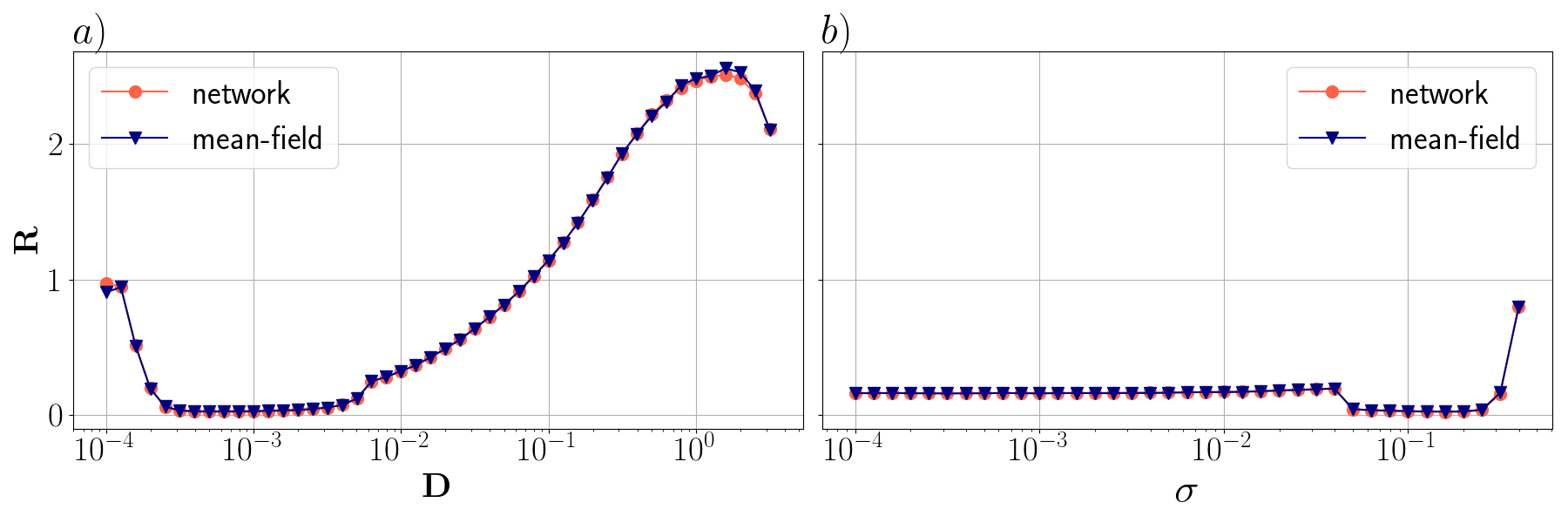}
	\caption{Normalized standard deviation of the interspike interval $R$ for a globally coupled network (red circles) with noise terms in both system variable equations and its mean-field system (blue triangles) with $a=1.05$: a) for fixed coupling strength $\sigma=0.1$ and varying noise intensity $D$, b) for fixed noise intensity \textbf{$D=0.001$} and varying coupling strength $\sigma$. The results are obtained by integrating over $10000$ time units and then averaging over time, oscillators, and realizations (5 simulations for each $\sigma$). The $x$-axis has logarithmic scaling. Other parameters: $N=100$, $\varepsilon=0.01$.}
	\label{fig:global_a_105}
\end{figure*}

\begin{figure*}
	\includegraphics[width=\textwidth]{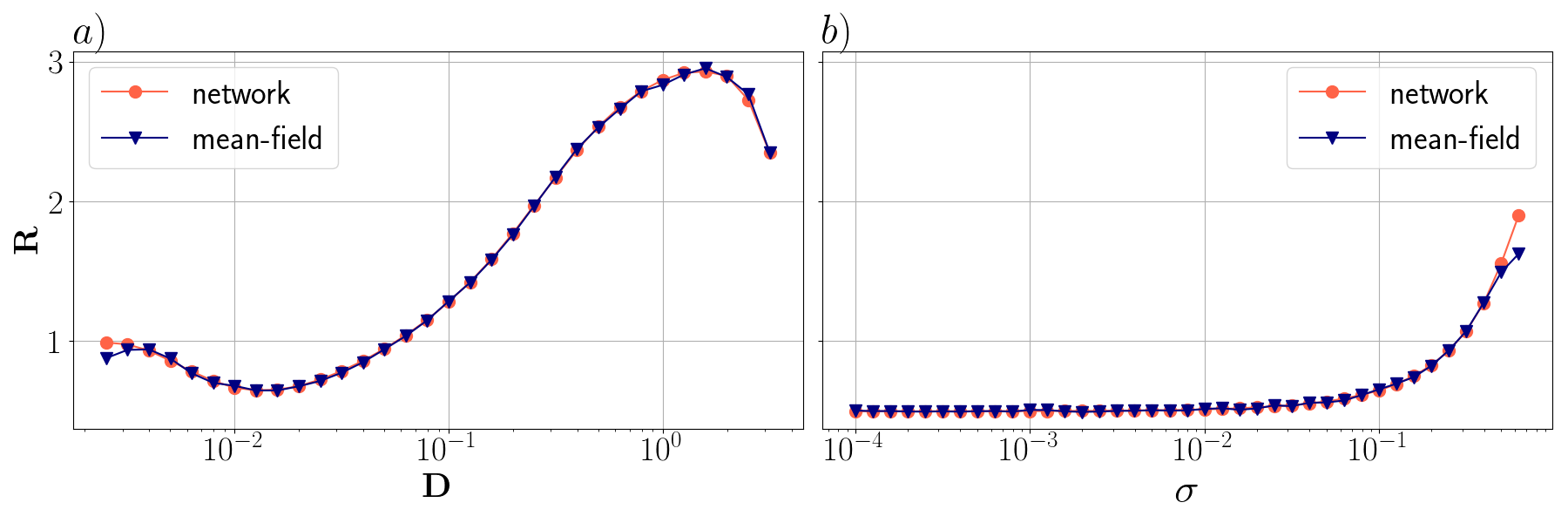}
	\caption{Normalized standard deviation of the interspike interval $R$ for a globally coupled network (red circles) with noise terms in both system variable equations and its mean-field system (blue triangles) with $a=1.3$: a) for fixed coupling strength $\sigma=0.1$ and varying noise intensity $D$, b) for fixed noise intensity \textbf{$D=0.0125893$} and varying coupling strength $\sigma$. The results are obtained by integrating over $10000$ time units and then averaging over time, oscillators, and realizations (5 simulations for each $\sigma$). The $x$-axis has logarithmic scaling. Other parameters: $N=100$, $\varepsilon=0.01$.}
	\label{fig:global_a_13}
\end{figure*}

Here we study the role of noise intensity $D$ and coupling strength $\sigma$ in inducing coherence resonance in a network of globally coupled FitzHugh-Nagumo oscillators. In particular, for the mean-field model we initially simulate the system given by Eqs. \eqref{eq:stochFHN}, by employing the classical Euler-Maruyama numerical scheme, as detailed in Appendix A. 
We measure $R$ in two different parameter settings: first we increase $D$, keeping all parameters fixed and second we increase $\sigma$, keeping all the other parameters fixed. The results
are shown in Figs. \ref{fig:global_a_105}, \ref{fig:global_a_13} for different excitability threshold values. Note that the $x$-axis is logarithmic in both cases.

Coherence resonance is visible both for $a=1.05$ (Fig.~\ref{fig:global_a_105} a)) and for $a=1.3$ (Fig.~\ref{fig:global_a_13} a)), where a minimum in $R$ emerges; the location of the minimum depends on the excitability threshold value and it occurs for different noise intensities $D$ in the two cases. It is worth noticing here that, if the system is closer to the Hopf bifurcation point, i.e., for $a=1.05$, it requires lower noise intensity for coherence resonance to occur. On the other hand, if the system is further away from the Hopf bifurcation point, i.e., for $a=1.3$, the system requires higher noise intensity. An interesting observation is that for both $a=1.05$ and $a=1.3$ the $R(D)$-curve has both minimum and maximum (Fig.~\ref{fig:global_a_105} a) and Fig.~\ref{fig:global_a_13} a)). The occurance of the maximum is associated with the phenomenon of anticoherence resonance investigated in \cite{LAC02,lindner2002maximizing,luccioli2006dynamical}. Here we show that the anticoherence is captured by the mean-field analysis.

To study the effects of coupling strength on the above observed coherence resonance, we measure $R$ as $\sigma$ is varied, for fixed $D$, in two different parameter settings: first, for $a=1.05$ (Fig. \ref{fig:global_a_105} b)) and second, for $a=1.3$ (Fig.~\ref{fig:global_a_13} b)). We observe for the case $a=1.05$ that coherence resonance is enhanced for a certain range of coupling strength (when $0.05 \leq \sigma < 0.25$). Several other works have also shown that coherence resonance can be enhanced by choosing appropriate coupling strengths \cite{HU00,KWO02,BAL14}. Here we demonstrate that this feature can be captured well in the mean-field framework. For the higher value of excitability threshold, i. e. $a=1.3$, the same can be observed. As can be seen in Fig. \ref{fig:global_a_13} b) the network behavior is captured well in the mean-field framework.

\begin{figure*}
	\includegraphics[scale=0.4]{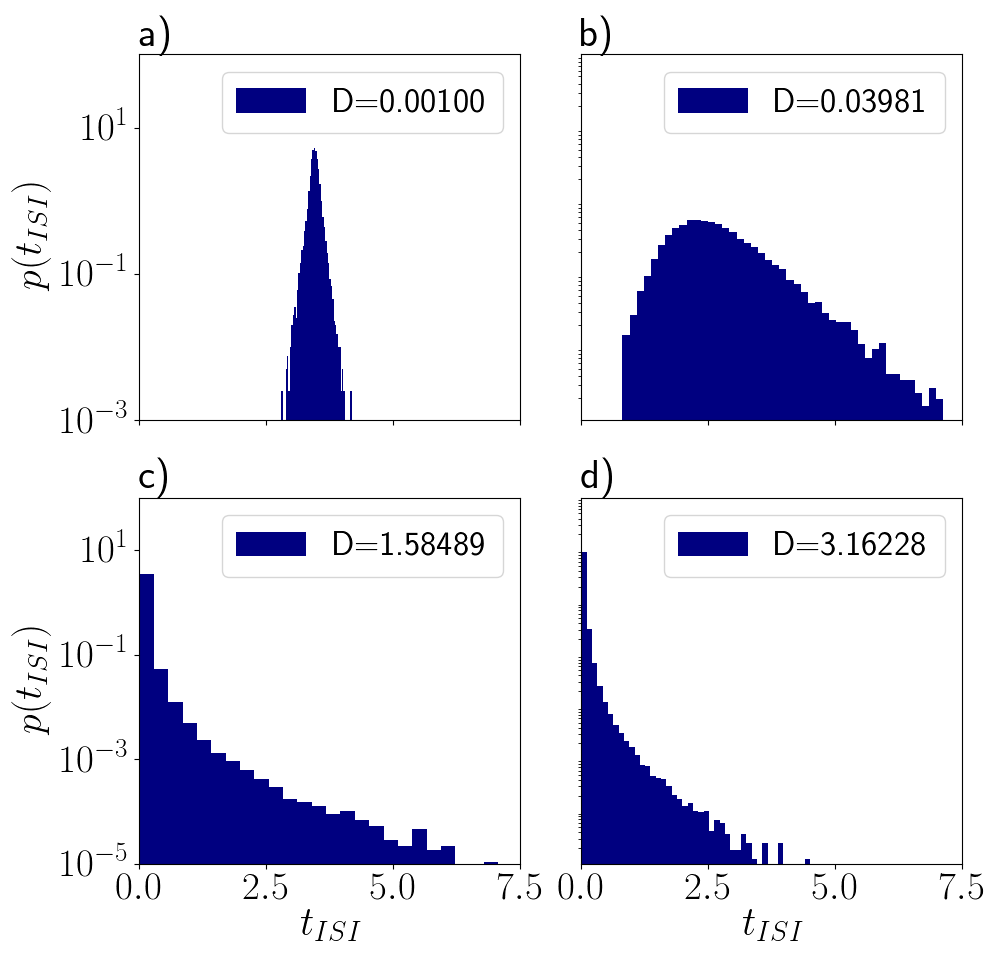}
	\caption{Probability distribution $p(t_{ISI})$ of the measured interspike intervals $t_{ISI}$ in a globally coupled network with $a=1.05$ and with noise terms in both system variable equations for noise intensity a) $D=0.001$ b) $D=0.03981$ c) $D=1.58489$ d) $D=3.16228$.}
	\label{fig:prob_a_105}
\end{figure*}

\begin{figure*}
	\includegraphics[scale=0.4]{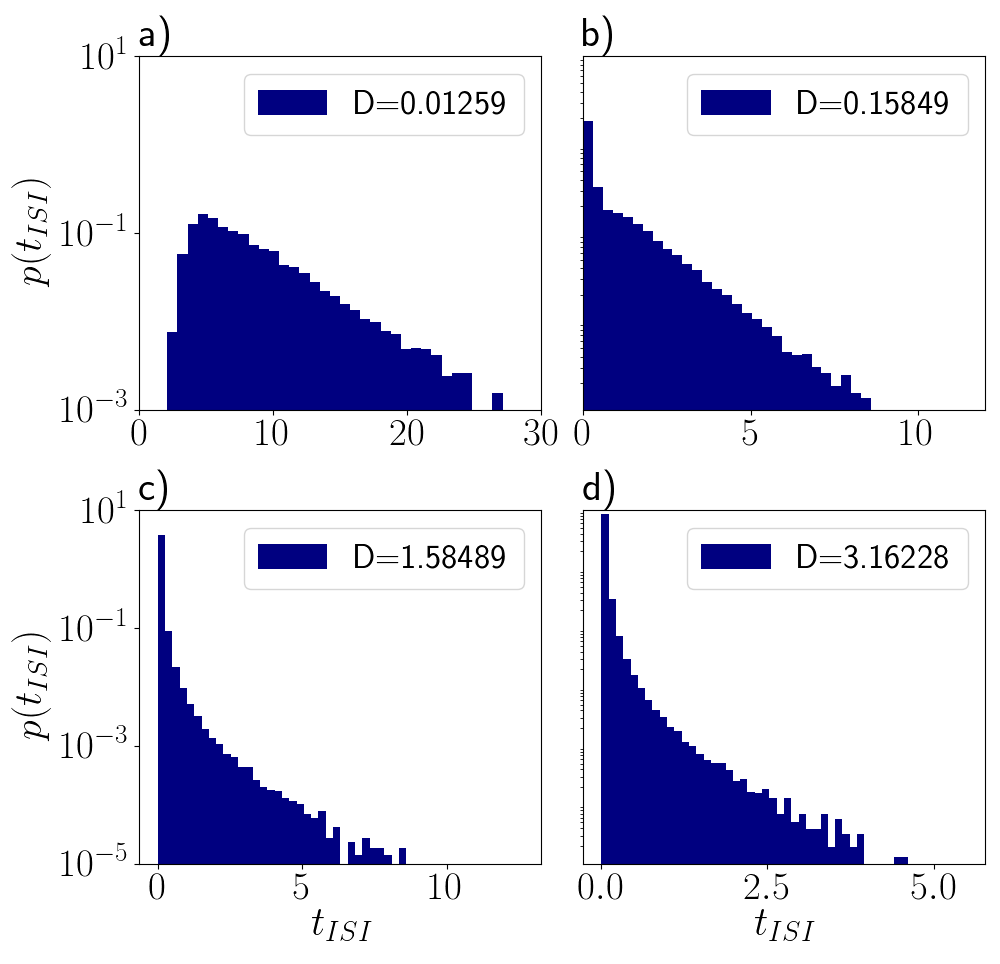}
	\caption{Probability distributions $p(t_{ISI})$ of the measured interspike intervals $t_{ISI}$ in a globally coupled network with $a=1.3$ and with noise terms in both system variable equations for noise intensity a) $D=0.01259$, which corresponds to the minimum of the curve shown in Fig.~\ref{fig:global_a_13} a); \\ b) $D=0.15849$; c) $D=1.58489$ corresponding to the maximum in Fig.~\ref{fig:global_a_13} a); d) $D=3.16228$.}
	\label{fig:prob_a_13}
\end{figure*}

The emergence of anticoherence resonance is due to the increasing role played by the noise, which destroys the refractory time proper of each neuron, thus allowing for infinitely small ISIs. This can be seen by plotting the probability distribution $p(t_{ISI})$ of the time between two successive spikes $t_{ISI}$, for increasing noise intensity (see Figs. \ref{fig:prob_a_105} and \ref{fig:prob_a_13} for $a=1.05$ and $a=1.3$ respectively). In particular in Fig. \ref{fig:prob_a_105} are reported the probability distributions for $D=0.001$ (panel a),  $D=0.03981$ (panel b), $D=1.58489$ (panel c) and $D=3.16228$ (panel d), thus characterizing four different states of the curve shown in Fig. \ref{fig:global_a_105} a), including the minimum and the maximum of the curve. At the minimum ($D=0.001$), the probability distribution is very peaked: the spike emission is coherent and all neurons spike with almost the same time interval. For increasing noise intensity ($D=0.03981$), the distribution becomes wider and longer times are possible between successive spikes. Finally at the maximum ($D=1.58489$) and for wider noise intensities, noise guides the dynamics and destroys the coherence: the presence of infinitely small ISIs is the signature of a fluctuation-driven dynamics, where neurons overcome the threshold very often due to the high noise intensity.  

Similar conclusions can be drawn from Fig. \ref{fig:prob_a_13}, that describes four different states of the curve previously reported in Fig. \ref{fig:global_a_13} a), corresponding to $D=0.01259$ (panel a), $D= 0.15849$ (panel b), $D= 1.5849$ (panel c) and $D=3.16228$ (panel d). Since  the system requires higher noise intensity to achieve coherence resonance at $a=1.3$, the level of coherence is lower with respect to the previous case and the probability distribution at the minimum ($D=0.01259$) is less peaked than the correponding case shown in Fig. \ref{fig:prob_a_105} a). For higher noise intensities (panels b-d), it is possible to register $t_{ISI}$ values that are smaller than the refractory time due to the presence of strong fluctuations that guide the network dynamics, thus triggering the anticoherence phenomenon.

The shown results are stable with respect to the integration scheme and do not depend on the chosen network size. More details are given in the Appendix A for the stability of the mean-field solution with respect to the integration scheme, while the impact of network size on the results is further discussed in the Appendix B. In the following we consider the globally coupled mean-field system with only one noise term, as given in Eqs. \eqref{eq:globallyCopuledMF}, which is the exact mean-field limit in the globally coupled case of Eqs. \eqref{eq:stochNetwork}. As previously done for the extended mean-field model, we study the role of noise intensity $D$ and coupling strength $\sigma$ in inducing coherence resonance and compare the results obtained from the mean-field and network simulations.

\begin{figure*}
	\includegraphics[width=\textwidth]{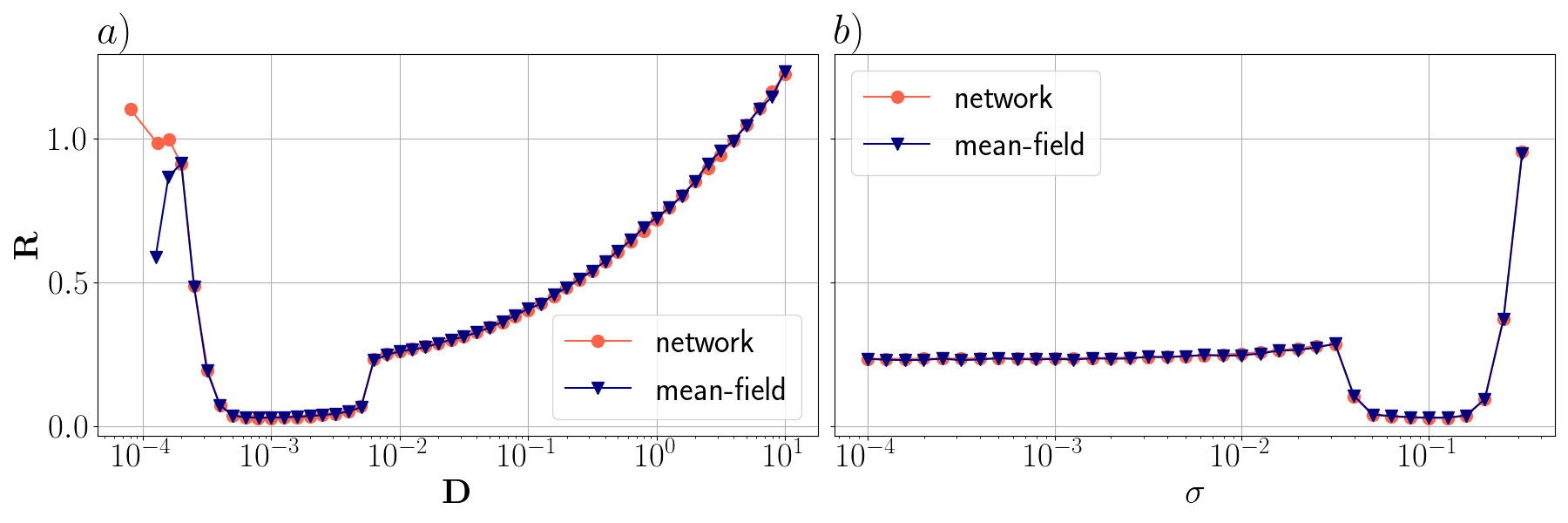}
	\caption{Normalized standard deviation of the interspike interval $R$ for a globally coupled network with single noise term as given in Eqs. \eqref{eq:stochNetwork} (red circles) and its mean-field system (blue triangles) with $a=1.05$: a) for fixed coupling strength $\sigma=0.1$ and varying noise intensity $D$ and b) for fixed noise intensity \textbf{$D= 0.00079$} and varying coupling strength $\sigma$. The results are obtained by integrating over $10000$ time units and then averaging over time, oscillators, and realizations (for 5 simulations each). The $x$-axis has logarithmic scaling. Other parameters: $N=100$, $\varepsilon=0.01$.}
	\label{fig:global_one_a_105}
\end{figure*}

\begin{figure*}
	\includegraphics[width=\textwidth]{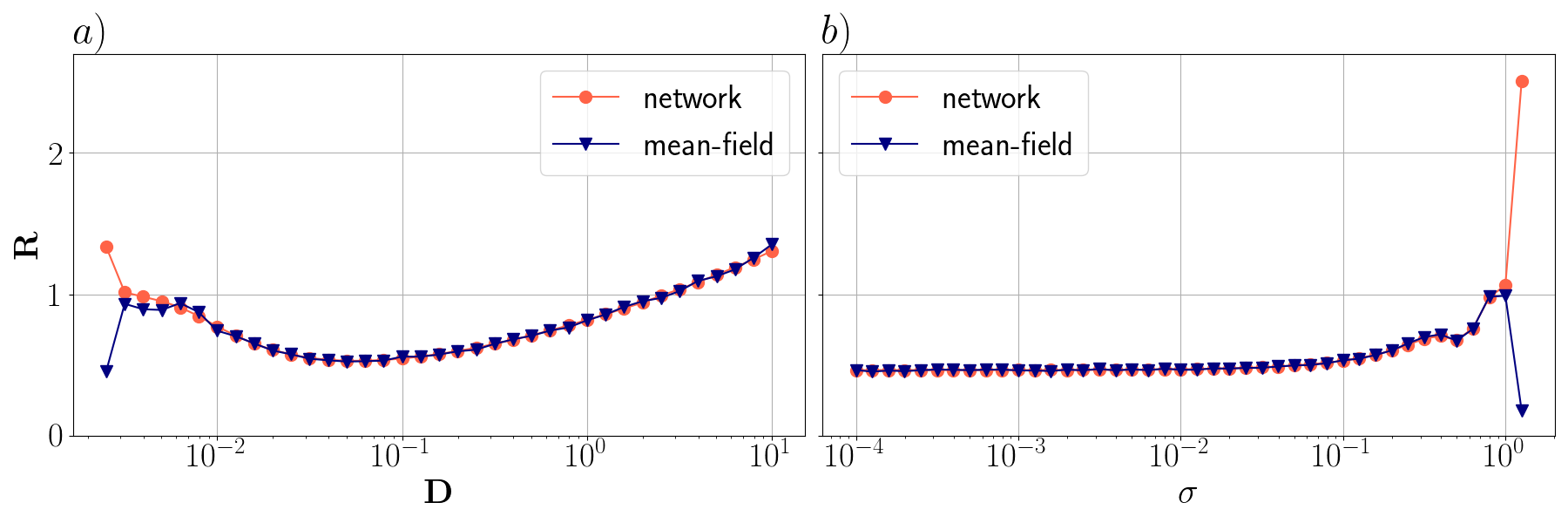}
	\caption{Normalized standard deviation of the interspike interval $R$ for a globally coupled network (red circles) with single noise term as given in Eqs. \eqref{eq:stochNetwork} and its mean-field system (blue triangles) with $a=1.3$: a) for fixed coupling strength $\sigma=0.1$ and varying noise intensity $D$ and b) for fixed noise intensity \textbf{$D= 0.05012$} and varying coupling strength $\sigma$. The results are obtained by integrating over $10000$ time units and then averaging over time, oscillators, and realizations (for 5 simulations each). The $x$-axis has logarithmic scaling. Other parameters: $N=100$, $\varepsilon=0.01$.}
	\label{fig:global_one_a_13}
\end{figure*}

In Figs.~\ref{fig:global_one_a_105} and ~\ref{fig:global_one_a_13} are shown the results for $a=1.05$ and $a=1.3$, respectively. In both cases coherence resonance is visible and a minimum in $R$ emerges (panel a).
As for the extended mean-field model, the location of the minimum depends on the excitability threshold value and it occurs for higher noise intensities $D$ when a higher excitability threshold is chosen.
Moreover, we observe that $R$ values obtained from the mean-field framework and the direct network simulation overlap almost completely for the whole considered range of noise values.  
In order to study the effects of coupling strength on the observed coherence resonance we have measured $R$ as $\sigma$ is varied for fixed $D$ (panel b): for both excitability threshold values coherence resonance can be obtained  by choosing appropriate coupling strengths. Also for this numerical experiment we see a good agreement between the network and the mean-field system.

Finally we observe a different behavior, in terms of anticoherence resonance, for the cases shown in Figs.~\ref{fig:global_one_a_105} a) and ~\ref{fig:global_one_a_13} a) with respect to the cases shown in Figs. \ref{fig:global_a_105} a) and \ref{fig:global_a_13} a): differently from the extended model, in the latter case we do not observe the presence of a relative maximum for high noise intensity values. 
A maximum appears in the mean-field system, for very small values of noise intensity, similarly to what reported in \cite{LAC02}, where anticoherence resonance has been shown to appear for a single stochastic FHN neuron, due to the mixing of two different time-scales: when adding additive noise of small intensity the neuron responds in what appears to be trains consisting of a few number of enchained pulses. While the pulses belonging to a single spike train are separated by a small deterministic time scale separation, the time interval between two consecutive pulses is longer and unpredictable. However the same conjecture cannot be extended to our globally coupled network of FHN neurons since we observe, in the network system, a purely Poissonian statistics, 
with the Poisson limit $R=1$ reached from the above. Therefore we conclude that the maximum in the mean-field system is rather an artificial effect due to the poor prediction of the model in the Poissonian regime.

\subsection{Locally coupled framework}

In the following we show the results obtained from the investigation of the mean-field system in the locally coupled regime (Eqs. \eqref{eq:FHN_local_coupled}), which has been simulated by applying the  
discrete scheme given in Eqs. \eqref{eq:FHN_local_discrete_notation}. These results are compared with the simulations of the network system (Eqs. \eqref{eq:stochNetwork}) with the equivalent topological structure.
Similarly to what shown in the globally coupled framework, we perform two different numerical experiments. In the first one, we keep the coupling strength parameter $\sigma$ fixed and vary the noise intensity level for two different excitability threshold values (i.e. $a=1.05$ and $a=1.3$). 
In the second experiment, we keep the noise intensity fixed and vary the coupling strength $\sigma$. The first set of experiments shows the relation between the mean-field approximation and the small noise requirement of the approximation. In the second set of experiments, we see the effect of the distance of the excitable system from the equilibrium point on the mean-field approximation.

\begin{figure}
	\centerline{\includegraphics[width=\textwidth]{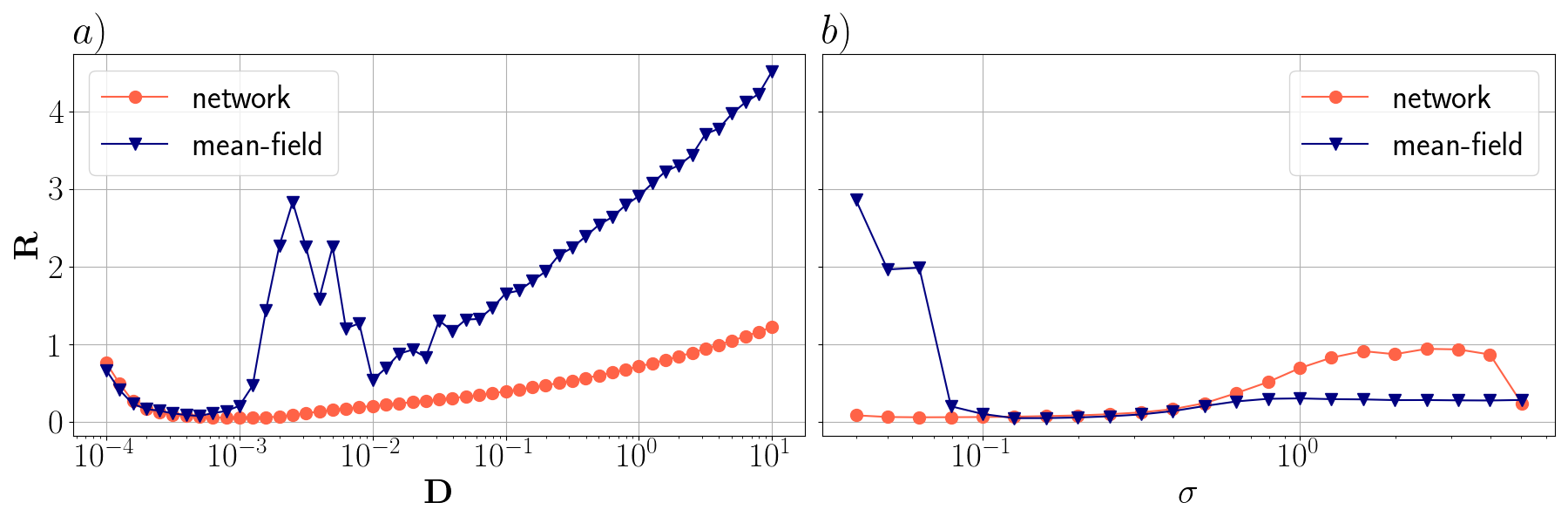}}
	\caption{Normalized standard deviations of the interspike interval $R$ for a locally coupled network (red circles) and the mean-field system (blue triangles) with $a=1.05$. Left: fixed coupling strength $\sigma=0.1$ and varying noise intensity $D$. Right: fixed noise intensity $D=0.0005$ and varying coupling strength $\sigma$. The results are obtained by integrating over 10000 time units and then averaging over time, oscillators, and realizations (for 5 simulations each). The  $x$-axis has logarithmic scaling. Other parameters: $N=100$, $\varepsilon=0.01$, $\Delta t=0.001$, $ T=M\Delta t=10000$. }
	\label{fig:local_a_105}
\end{figure}

\begin{figure}
	\centerline{\includegraphics[width=\textwidth]{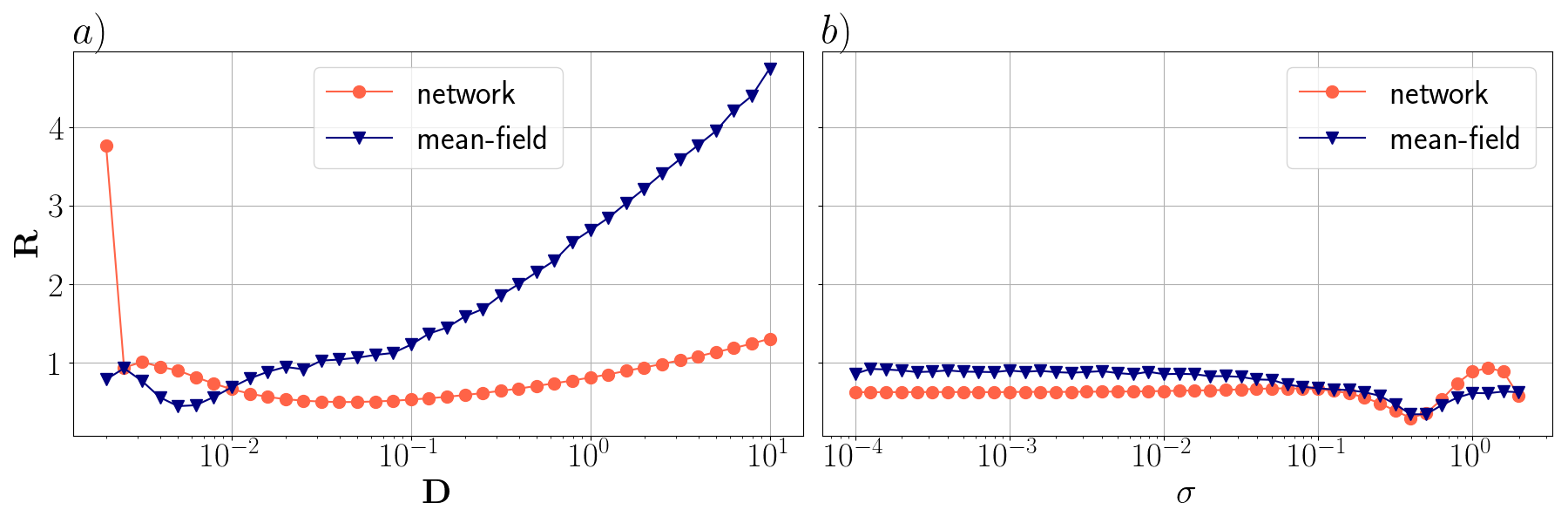}}
	\caption{Normalized standard deviations of the interspike interval $R$ for a locally coupled network (red circles) and the mean-field system (blue triangles) with $a=1.3$. Left: fixed coupling strength $\sigma=0.1$ and varying noise intensity $D$. Right: fixed noise intensity $D=0.001$ and varying coupling strength $\sigma$. The results are obtained by integrating over 10000 time units and then averaging over time, oscillators, and realizations (for 5 simulations each). The  $x$-axis has logarithmic scaling. Other parameters: $N=100$, $\varepsilon=0.01$, $\Delta t=0.001$, $\quad T=M\Delta t=10000$. }
	\label{fig:local_a_13}
\end{figure}

The dependence on the noise intensity of the normalized standard deviation $R$ of the interspike intervals is shown in Fig. \ref{fig:local_a_105} for $a=1.05$ and in Fig. \ref{fig:local_a_13} for $a=1.3$, respectively.
While the parameters $\varepsilon=0.01$, $\sigma=0.1$ have been kept fixed, the noise intensity level $D$ has been varied over logarithmically sampled values between $0.0001$ and $10$ for $a=1.05$ on the left panel. It can be seen that $R$ depends non-monotonically on $D$ and a minimum of this quantity is observable for small noise intensity, thus indicating coherence resonance. 
We remark that in the case of $a=1.05$, for the mean-field system, the coherence of oscillations grows ($R$ decreases) as we increase the noise intensity $D$ towards the value $6.30957\times 10^{-4}$, where it reaches its highest level, as shown in Fig. \ref{fig:local_a_105} a) in terms of blue triangles. Then it starts decaying as we increase $D$ further. 
The coherence indicator $R$ shows, for the network system, a pattern similar to the one obtained in the mean-field framework until the noise values are close to and smaller than $10^{-3}$, as shown in Fig. \ref{fig:local_a_105} in terms of red circles. As we increase the noise intensity $D$ towards values higher than $6.30957\times 10^{-4}$, the coherence of the oscillations starts decreasing and the mean-field curve does not overlap with the one obtained from the network equations any longer. This is due to the fact that we approximate the state variables $u$ and $v$ via Gaussian random processes for small noise intensity values to derive the mean-field system and this approximation does not hold for high noise intensity values (i.e. $D>6.30957\times 10^{-4}$). On the right panel, we keep the level of noise intensity fixed and vary the coupling strength. We observe that the mean-field model overlaps with the network in the coherence resonance region, while there is no overlap outside that region.

In the case of $a=1.3$, the excitable system is further away from the border between excitable and oscillatory regimes in comparison to the case of $a=1.05$. This results in a rather incoherent spiking of neurons and the Gaussian approximation is not valid, thus affecting the mean-field predictions, as can be seen in Fig. \ref{fig:local_a_13} a). On the right panel, where the coupling strength is varied while keeping the noise intensity level fixed, we observe that the mean-field model and the network coincides in the coherence resonance region as in the case with $a=1.05$. In the rest, although they follow the same pattern (almost constant) for small coupling strength values, they do not overlap outside the coherence resonance region.

We note that, in both cases $a=1.05$ and $a=1.3$, the mean-field is able to reproduce the network dynamics only when the network spikes coherently. In this regime the variance is low and it is possible to observe a regular spike activity in the mean-field system. For small coupling strength values, while there are always some neurons in the network which spike occasionally, it is not possible to observe spike emissions in the mean-field system due to the high variance values at each time instant: due to the high variance, the mean-field system drags $u$ to negative values as soon as $u$ exceeds $0$, thus impeding the spike emission. This is at the origin of the different behaviors observable at $a=1.05$ for small coupling strength values (see Fig. \ref{fig:local_a_105} b)).
At $a=1.3$ the network is further away from the equilibrium point, therefore less neurons spike at small coupling strength values, with respect to the previous case.
It results in lower variance values that explains the better agreement with the mean-field system Eqs.~\eqref{eq:FHN_local_coupled}, since a comparable amount of spikes can be registered in both systems 
(see Fig. \ref{fig:local_a_13} b)).

\section{Conclusion}
\label{sec:conclusio}
In this paper we have developed a mean-field framework that allows analytical treatment of coherence resonance in complex networks of FHNs. In particular we have compared the obtained results with the direct numerical simulation of the network equations for the locally and globally coupled networks.

For the globally coupled case, we demonstrate a good agreement. Moreover, all the ``nuances'' of coherence resonance, such as sensitivity to excitability threshold and the coupling strength, are captured by the mean-field framework. On the other hand, for the locally coupled case, we have disagreement for large noise values, due to the fact that we approximate the state variables via Gaussian random processes for small noise intensity values. Therefore, we identify the limitations of our mean-field framework that works well only for small noise values in the locally coupled case.

The better agreement for the globally coupled case compared to the locally coupled case can be explained by the fact that the two mean-field models are different for the two cases. The locally coupled case needs the variances $\rho^2$ of $u$ at each time step. Since the excitation travels through the network (compare Fig. \ref{fig:local_a_13}), the variances become quite large when spikes occur, making it harder, if not impossible, to capture the full dynamics in a mean-field model. The globally coupled case simply takes the mean of all oscillators. As all oscillators have the same coupling term, they also show the same spiking behavior. This high similarity in the behavior leads to a very small variance in $u$ (and $v$), meaning that the mean-field captures the entire behavior.
Further, the better agreement for the globally coupled case compared to the local topology can be explained by the push-pull effect generated by the all-to-all interaction among the neurons. This allows neurons to spike coherently, while in the locally coupled architecture, due to the absence of such push-pull effect, highly varied spiking patterns emerge for each neuron in the network.
This affects the mean-field model via the $\rho$ term, thus preventing the mean-field from giving reasonable insights in the network dynamics (see Eqs. \eqref{eq:FHN_local_coupled}).

Finally in the globally coupled system we have found, for the extended model Eqs. \eqref{eq:stochFHN}, anticoherence resonance, which takes place for high noise intensity values. 
For the extended model anticoherence is originated from noise induced activation processes: the noise is so strong that guides neurons over threshold continuously, thus inducing firing emissions at infinitely small $t_{ISI}$ values. At the contrary, for the original mean-field model Eqs. \eqref{eq:globallyCopuledMF}, the anticoherence phenomenon cannot be observed neither at low nor at high intensity values.

Interesting future research directions on the topic would be to extend the mean-field framework to a multi-layer topology, where the emergence and control of coherence resonance have been recently found \cite{SEM18, yamakou2019control} and to introduce biological features in the model, like electrical and chemical synapses, that play functional roles in information processing, similarly to what shown in \cite{yamakou2020optimal} for self-induced stochastic resonance.

\begin{acknowledgments}
This work was supported by the Deutsche Akademische Austauschdienst (DAAD, German Academic Exchange Service) -  Projektkennziffer - 57445304 - PPP Frankreich Phase I and the Deutsche Forschungsgemeinschaft (DFG, German Research Foundation) - Projektnummer - 163436311 - SFB 910.
From the French side, this work was supported by Campus France - programme PHC PROCOPE 2019 - Num\'{e}ro de projet : 42511TA.
\end{acknowledgments}

\FloatBarrier

\section*{Appendix A: Integration scheme}

In the globally coupled framework we simulate the system given by Eqs. \eqref{eq:stochFHN}, by employing the classical Euler-Maruyama numerical scheme.  The discretized equations read as
\begin{equation}\label{eq:FHN_global_discrete_notation}
\begin{split}
\hat{u}(t_{k+1})-\hat{u}(t_{k})= & \frac{\Delta t}{\eps}\Big\{f\Big(\hat{u}, \hat{v} \Big) + \sigma\Big(\mathbb{E}[\hat{u}(t_k)] - \hat{u}(t_k) \Big)\Big\}+ \sqrt{\frac{2 \bar{D}}{\eps}}d\hat{\bar{W}}       ,\\
\hat{v}(t_{k+1})-\hat{v}(t_{k})= & \Big(\hat{u}(t_k)+a\Big)\Delta t+\sqrt{2D}\,d\hat{W}(t_k),
\end{split}
\end{equation}
where the $\hat{}$ denotes the discretized variables in time, $\Delta t$ is the time step and $t_k$ is the $k^{\text{th}}$ sample of time. More precisely  $t_k=k\Delta t$ with $k=0,1,\dots, M$, where the initial and final time instants of the  evolution are $t_0=0$ and $t_M=T=M\Delta t$, respectively. 

In the Euler-Maruyama scheme the noise terms are discretized as
\begin{equation}
d\tilde{\bar{W}}(t_k)=\sqrt{\Delta t}\,\bar{\mu}(t_k), \quad d\tilde{W}(t_k)=\sqrt{\Delta t}\,\mu(t_k),
\end{equation}
where $\bar{\mu}(t_k)$ and $\mu(t_k)$ are random numbers generated independently from the standard normal distribution at each time instant $t_k$.

In the locally coupled framework we simulate the system given by Eqs. \eqref{eq:FHN_local_coupled}. We employ the classical Euler-Maruyama numerical scheme where we use the following explicit discretized equations:
\begin{equation}\label{eq:FHN_local_discrete_notation}
\begin{split}
\hat{u}(t_{k+1})-\hat{u}(t_{k})= & \frac{\Delta t}{\eps}\Big\{ \Big( 1-\hat{\rho}^2(t_k) \Big)\hat{u}(t_k)-\frac{\hat{u}^3(t_k)}{3} -\hat{v}(t_k)     \Big\},\\
\hat{v}(t_{k+1})-\hat{v}(t_{k})= & \Big(\hat{u}(t_k)+a\Big)\Delta t+\sqrt{2D}\,d\hat{W}(t_k),
\end{split}
\end{equation}
where the notation and the parameters are the same as in Eqs. \eqref{eq:FHN_global_discrete_notation}.
In order to perform the numerical integration, the value of $\mathbb{E}[\hat{u}]$ (respectively $\hat{\rho}^2$) is needed at each time instant $t_k$ for the globally (locally) coupled setting. 
We find the values of $\mathbb{E}[\hat{u}]$ (respectively $\hat{\rho}^2$), at each time step, by integrating the network Eqs. \eqref{eq:stochNetwork}, with initial conditions generated from a Gaussian distribution and by implementing the Euler-Maruyama method.
In particular we have performed $L=20$ simulations of globally (locally) coupled networks, where the initial conditions are randomly generated in each simulation. This provides $L$ different set of $\mathbb{E}[\hat{u}]$ (respectively $\hat{\rho}^2$) values for each time instant. 
Finally the  $\mathbb{E}[\hat{u}]$ (respectively $\hat{\rho}^2$) values to be introduced in the associated mean-field framework, given by  Eqs. \eqref{eq:FHN_global_discrete_notation} (Eqs. \eqref{eq:FHN_local_discrete_notation}), are determined as the average over $L$ of the values calculated for each time instant. 

\begin{figure}
	\includegraphics[width=0.5\textwidth]{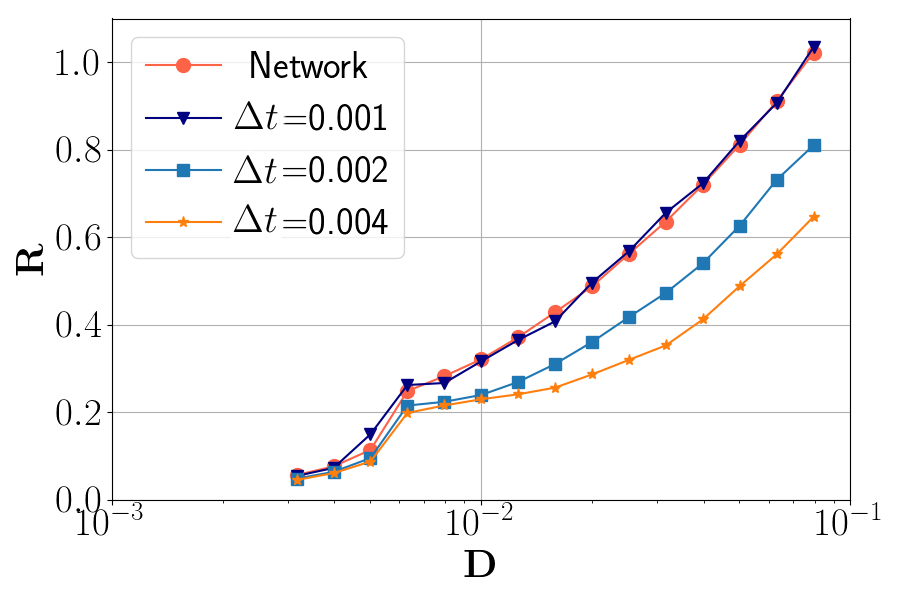}
	\caption{Normalized standard deviation of the interspike interval $R$ for a globally coupled network (red circles) with $N=100$ nodes and with noise terms in both state variable equations; and the same of the corresponding mean-field system simulated with different time steps: $\Delta t=0.001$ (dark blue triangles), $\Delta t=0.002$ (light blue squares) and $\Delta t=0.004$ (orange stars) for fixed coupling strength $\sigma=0.1$ and varying noise intensity $D$. The results are obtained by integrating over $5000$ time units and then averaging over time and oscillators. The $x$-axis has logarithmic scaling. Other parameters: $a=1.05$, $\varepsilon=0.01$.}
	\label{fig:stepsize}
\end{figure}

Finally, we investigate the influence of the time step size used for the integration of the globally coupled mean-field  Eqs. \eqref{eq:stochFHN}. It is important to use a small time step ($\Delta t=0.001$) to achieve good agreement of mean-field results with those obtained from numerical simulation of the network equations (see Fig. \ref{fig:stepsize}). Using a larger time step leads to the disagreement of the results for strong noise.

\section*{Appendix B: Dependence on the network size}

\begin{figure*}
	\includegraphics[width=0.5\textwidth]{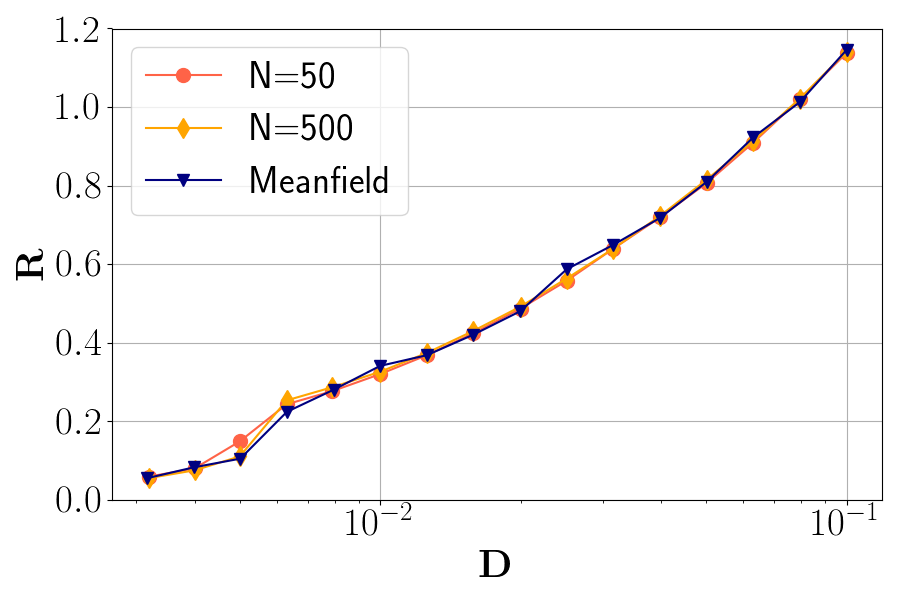}
	\caption{Normalized standard deviations of the interspike interval $R$ for a globally coupled network (red circles) with noise terms in both system variable equations. Cases with $N=50$ (red circles) and $N=500$ nodes (orange diamonds) are compared to the mean-field system (blue triangles) where coupling strength $\sigma$ is fixed to $0.1$ and noise intensity $D$ is varied. The results are obtained by integrating over $5000$ time units and then averaging over time and oscillators. The $x$-axis has logarithmic scaling. Other parameters: $a=1.05$, $\varepsilon=0.01$.}
	\label{fig:finite_size}
\end{figure*}

Here we analyze the impact of network size on our results. In more detail, we compare $R(D)$ curves obtained for the globally coupled case from the direct numerical simulation of the network equations for $N=50$ and $N=500$ with the curve resulting from the mean-field framework. In particular, for the mean-field model we simulate the system given by Eqs. \eqref{eq:stochFHN} and analogously, for the network equations, we integrate Eqs. \eqref{eq:stochNetwork} with noise terms in both system variable equations. It turns out that the system size does not have impact here and all three curves agree well (see Fig. \ref{fig:finite_size}). 

%

\end{document}